\documentclass[letterpaper, 10 pt, conference]{ieeeconf}  %

\IEEEoverridecommandlockouts                              %

\usepackage{graphics} %
\usepackage{epsfig} %
\usepackage{times} %
\usepackage{amsfonts} %
\usepackage{amsmath}%
\usepackage{amssymb}%
\usepackage{mathtools}%
\newcommand{\norm}[1]{\left\lVert#1\right\rVert} %
 
\newtheorem{theorem}{Theorem}

\usepackage{algorithm}%
\usepackage{algpseudocode} %
\usepackage{xspace}%
\newcommand{\eg}{e.g.~}%
\newcommand{\ie}{i.\,e.\xspace}%
\usepackage{siunitx}
\usepackage[short,nocomma]{optidef} %
\usepackage{hyperref}%

\DeclareMathOperator*{\argmaxA}{arg\,max}

\usepackage{pgfplots}
\usepackage{tikz}
\pgfplotsset{compat=newest}
\usetikzlibrary{plotmarks}
\usetikzlibrary{arrows.meta}
\usepgfplotslibrary{patchplots}

\newcommand\submittedtext{%
  \footnotesize \textcopyright \the\year{} IEEE. Personal use of this material is permitted. Permission from IEEE must be obtained for all other uses, including reprinting/republishing this material for advertising or promotional purposes, collecting new collected works for resale or redistribution to servers or lists, or reuse of any copyrighted component of this work in other works.}

\newcommand\submittednotice{%
\begin{tikzpicture}[remember picture,overlay]
\node[anchor=south,yshift=10pt] at (current page.south) {\fbox{\parbox{\dimexpr0.65\textwidth-\fboxsep-\fboxrule\relax}{\submittedtext}}};
\end{tikzpicture}%
}
\title{\LARGE \bf
Safe and High-Performance Learning of Model Predictive Control using Kernel-Based Interpolation}

\author{Alexander Rose$^1$, Philipp Schaub$^{1}$, Rolf Findeisen$^1$%
\thanks{$^1$Control and Cyber-Physical Systems Laboratory, Technical University of Darmstadt, Germany, {\small \{alexander.rose, pschaub, rolf.findeisen\}@iat.tu-darmstadt.de}}
}

\begin{document}

\maketitle
\thispagestyle{empty}
\pagestyle{empty}

\begin{abstract}

We present a method that allows efficient and safe approximation of model predictive controllers using kernel interpolation.
Since the computational complexity of the approximating function scales linearly with the number of data points, we propose to use a scoring function which chooses the most promising data.
To further reduce the complexity of the approximation, we restrict our considerations to the set of closed-loop reachable states. 
That is, the approximating function only has to be accurate within this set.
This makes our method especially suited for systems, where the set of initial conditions is small.
In order to guarantee safety and high performance of the designed approximated controller, we use reachability analysis based on Monte Carlo methods.

\end{abstract}

\section{INTRODUCTION}

\submittednotice

Model predictive control (MPC) is an advanced control method, which can naturally handle nonlinear multi input multi output systems that are subject to constraints.
Moreover, MPC can be combined with machine learning methods to improve the control performance or predictive power of the model, while ensuring safe control of a dynamical system.
An overview of different approaches to integrating machine learning into MPC is presented in \cite{HewingLearningReview}. 
 Consequently, MPC is widely applied in industries ranging from pharmaceutical production to power electronics \cite{Schwenzer2021, QINMPCindustry}.

However, MPC requires solving a nonlinear programming problem in real time.
Consequently, the application of MPC in situations with limited computational resources, or where fast decisions are necessary, is challenging.
Therefore, various approaches exist to reduce the computational burden of MPC.
 For example, an efficient method to solve the nonlinear optimization problems is proposed in \cite{DiehlRTI}.
Another way to reduce the complexity of the optimization problem is for example by solving the problem in a lower dimensional subspace \cite{Schurig_dimreductionGrass, PANdimred}.

Instead of efficiently solving the optimization problem, another approach to alleviate the computational burden is to find a function which explicitly approximates the solution of the optimization problem.
To this end, the problem typically has to be solved prior to application for a lot of cases and thus most of the computational burden is put into an offline phase.
During the online employment no optimization problem has to be solved and only the approximated function has to be evaluated.

For example, in \cite{Bemporad_eMPC_complex, JOHANSEN2004293} the MPC law is approximated by multiple piece-wise affine functions which are sufficiently accurate in different regions of the state space.
Online computational complexity then boils down to finding the region for a given state and evaluate the affine function.
In \cite{Bayer_TubeExplicit2016}, a tube based MPC approach is used to partition the space into regions such that the MPC is robust to all possible initial states within each region.
A drawback of these approaches is that the number of regions drastically increases with the problem size limiting its application to small scale systems \cite{Bemporad_EMPC2009}.

To avoid partitioning the state space into multiple regions, an alternative way is  to use general nonlinear function approximators to learn the MPC law.
To this end, on the one hand, there are parametric regression methods and, on the other, there are non-parametric methods.
Parametric methods essentially store the available data by adjusting its model parameter.
A famous example of such methods are neural networks.
Some of the early works on approximating an MPC law using neural networks is for example presented in 
\cite{PARISINI1995}.
The authors in \cite{Karg2020} show that neural networks using the rectified linear unit transfer function can indeed exactly represent piecewise affine MPC laws.
In \cite{Nguyen21_NNController}, the authors provide a method to verify closed loop stability of neural network based MPC approximations. 
Instead of learning the MPC law directly,  in \cite{chatzikiriakos2024learning} the authors approximate the value function of the optimization problem. 
Consequently, to obtain the control input a much simpler optimization problem has to be solved online.
The advantage of this approach is that the authors can directly exploit properties of the value function to derive safety guarantees.

To our knowledge, less research exists on using non-parametric regression methods to approximate the MPC law.
Instead of storing the information in model parameters, these methods directly use the data to make predictions.
While this can lead to a high computational complexity if a lot of data are used in the model, there are several advantages compared to parametric methods.
When approximating an MPC law, the data are noiseless so that the task essentially becomes an interpolation task which can naturally be handled with non-parametric methods. %
Furthermore, one can keep the amount of needed data small, if they are selected in a reasonable way.
Overall, this can lead to a simplified design and verification of the approximated controller when using non-parametric methods.

For example, in \cite{SASAKI18_ExplicitMPC_GP_Flows} a Gaussian process based approximation of an MPC law is used to control the flow around a cylinder. 
While the authors do not provide rigorous guarantees, the controller shows good performance in simulation.
Another example is the work in \cite{GANGO2019152}, where the authors use a Gaussian process based approximation for linear parameter varying systems.
For linear systems, in   \cite{LygerosAMPCGP} an approximation using Gaussian processes is considered.
The authors propose a design procedure with probabilistic guarantees.
In \cite{tokmak2023automatic}, the authors propose a method for kernel based interpolation of an MPC law.
They provide guarantees on the maximum number of needed samples and error bounds of the approximation based on the theory of reproducing kernel Hilbert spaces.

Often the approximation methods require to first design a controller, which is robust to the approximation error, see \eg \cite{RoseLearnMPCGP, HertneckAMPC, tokmak2023automatic}. This can make the design rather challenging.

Instead, in this work we propose to use reachability analysis to validate the approximated controller with respect to safety and performance requirements. 
Thereby, we overcome the need to design a provably robust controller as long as the nominal controller has sufficient inherent robustness properties.
While similar approaches exist using neural networks, see \eg \cite{KochdumperNNReach, Karg2021reachandperformance}, we specifically propose a kernel-based interpolation method. 
This allows us to propose an efficient sampling strategy, leading to an overall small number of samples needed for the approximation and thus to low computational complexity while satisfying safety and performance requirements. %
 Our approach is specifically suited for cases, where the set of possible initial conditions is limited. 
In these cases only a small part of the MPC law has to be approximated accurately, since only a limited subset of the state space is reachable in closed loop.
This potentially allows to apply our method for high dimensional systems, while keeping the number of necessary data in the model small.

Note that in a more general context, not directly tied to MPC, the problem of finding a computationally favorable approximation of some model or data can also be viewed as an imitation learning \cite{OsaImitationLearningSurvey} or emulation \cite{GPemulator} problem. 

This paper is structured as follows.
We first present an overview of model predictive control followed by a brief introduction to reachability analysis. 
We then propose to use a Monte Carlo based method to approximately compute reachable sets, which allows us to derive probabilistic guarantees.
Thereafter, we present the main contribution of the paper. That is, we propose a method for high performance and safe approximation of a model predictive controller using kernel interpolation.
Finally, we illustrate our method in a case study.

\section{MODEL PREDICTIVE CONTROL}

We consider nonlinear discrete time systems
\begin{align}
    x(k+1) =  f(x(k),u(k)),
\end{align}
where the system dynamics is described by a nonlinear Lipschitz continuous function $f:\mathbb{R}^{n_x} \times \mathbb{R}^{n_u} \rightarrow  \mathbb{R}^{n_x}$. We denote  $x \in \mathbb{R}^{n_x}$ as the state, $u\in \mathbb{R}^{n_u}$ as the input of the system and $k \in \mathbb{N}_{\geq 0}$ as the discrete time index.
Furthermore, we define a sequence of states
of length $N_p+1$ as  $\{X\}_0^{N_p} = \{ x(0), \ldots, x(N_p) \}$ and a corresponding sequence of inputs as $\{U\}_0^{N_p-1} = \{u(0),\ldots,u(N_p-1)   \}$.
Herein, we call $N_p \in \mathbb N$ the prediction horizon.
The objective function assigns a real value to given sequences
\begin{align} \label{eq:objective}
    J(\{X\}^{N_p}_{0},\{U\}^{N_p-1}_{0}) = \sum_{i=0}^{N_p-1} \ell(X_i,U_i) + E(X_{N_p}),
\end{align}
evaluating their performance.

The goal in optimal control is to find sequences $\{X\}_0^{N_p}$ and $\{U\}_0^{N_p-1}$, which solve the following optimal control problem given an initial state $x$

\begin{mini}%
    {\{U\}_0^{N_p-1}}{J(\{X\}_0^{N_p},\{U\}_0^{N_p-1})}{\label{eq:ocp}}{ V(x) =}
    \addConstraint{\bar{x}(i+1)}{=f(\bar{x}(i),\bar{u}(i)), \, \bar{x}(0) = x}{,}
    \addConstraint{\bar{x}(i)}{\in \mathcal{X}_\text{con} \ominus \tilde{\mathcal{X}}, \, \bar{x}(N_p) \in \mathcal{X}_\text{term},}{}
    \addConstraint{\bar{u}(i)}{ \in \mathcal{U}_\text{con} \ominus \tilde{\mathcal{U}},}{}
    \addConstraint{\{U\}_0^{N_p-1}}{= \{ \bar{u}(0),\dots, \bar{u}(N_p-1) \},   }{}
    \addConstraint{\{X\}_0^{N_p}}{= \{ x, \bar{x}(1), \dots, \bar{x}(N_p) \}. }{}
\end{mini}

We denote the corresponding optimal sequences as $\{X^*(x)\}_0^{N_p}$ and $\{U^*(x)\}_0^{N_p-1}$.

In \eqref{eq:ocp} we use $\bar{(\cdot)}$ to emphasize a predicted value and $\ell \! : \! \mathbb{R}^{n_x}\times \mathbb{R}^{n_u} \rightarrow \mathbb{R}_{\geq 0} $ is a positive semi definite stage cost.%

Furthermore, $\mathcal{X}_\text{con}$ and $\mathcal{U}_\text{con}$ denote state and input constraints, respectively.
Finally, \eqref{eq:ocp} implicitly defines a model predictive controller
\begin{align} \label{eq:mpclaw}
\kappa(x) = U_0^*(x).%
\end{align}
In \eqref{eq:ocp}, we additionally tighten the constraints by $  \tilde{\mathcal{U}} $ and $\tilde{\mathcal{X}}$. This way an approximation of \eqref{eq:mpclaw} can make errors and still satisfy the demanded constraints $\mathcal{U}_\text{con}$,  $\mathcal{X}_\text{con}$ .

To guarantee stability and repeated feasibility of the system controlled by \eqref{eq:mpclaw}, we introduce terminal constraints $\mathcal{X}_\text{term}$ together with a suitable end penalty $E \! : \! \mathbb{R}^{n_x} \rightarrow \mathbb{R}_{\geq0}$, see \eg \cite{Findeisen2002,Gruene2017}.
We denote the set for which a solution to \eqref{eq:ocp} exists as $\mathcal{X}_\text{feas}$.

For simplicity, we additionally introduce the following notations.
Given a controller $\tilde{\kappa}: \mathbb{R}^{n_x} \rightarrow \mathbb{R}^{n_u}$,  we denote the controlled system as 
\begin{align}
    x(k+1) = f(x,\tilde{\kappa}(x)) = f_{\tilde{\kappa}}(x).
\end{align}

This way, we can for example express the closed loop state $j$ steps ahead of $k$ as
\begin{align}
    x(k+j) = \underbrace{f_{\tilde{\kappa}} \circ \ldots  
\circ f_{\tilde{\kappa}}(x(k))}_{j\, \text{times}}  = f_{\Tilde{\kappa}}^{\circ j }(x(k)).
\end{align}

To evaluate the closed loop performance of a given controller $\Tilde{\kappa}$, we generate closed loop trajectories with $N_\text{sim} \in \mathbb{N}$ steps, \ie we compute $\{X\}_0^{N_\text{sim}}=\left[x, \, f_{\tilde{\kappa}}(x), \dots, f_{\tilde{\kappa}}^{\circ N_\text{sim}}(x) \right]$ 
and $\{U\}_0^{N_\text{sim}-1} = \left[\tilde{\kappa}(x), \, \tilde{\kappa}(f_{\tilde{\kappa}}(x)), \dots, \tilde{\kappa}( f_{\tilde{\kappa}}^{\circ N_\text{sim}-1}(x)) \right]$.
The closed loop performance is then
\begin{align} \label{eq:cl_performance}
    P_{\tilde{\kappa}}(x) = \sum_{i=0}^{N_\text{sim}-1} \ell(X_i, U_i).
\end{align}
We can compare a controller $\Tilde{\kappa}$ with the implicitly defined predictive controller $\kappa$ by means of the absolute performance deviation
\begin{align}
     P_{\Tilde{\kappa},\text{abs}}(x) =  \norm{P_{\Tilde{\kappa}}(x) - P_\kappa(x)} 
\end{align}
and relative performance deviation
\begin{align}
    P_{\Tilde{\kappa},\text{rel}}(x) = \frac{  \norm{P_{\Tilde{\kappa}}(x) - P_\kappa(x)} }{P_{\kappa}(x)}.%
\end{align}
Notice that the implicitly defined predictive controller $\kappa$ in \eqref{eq:mpclaw} does not necessarily minimize \eqref{eq:cl_performance}, since $\kappa$ is obtained by repeatedly solving \eqref{eq:ocp} and in general predicted and closed loop trajectory differ.
In order to promote closed loop performance, we can choose  a longer prediction horizon $N_p$ in \eqref{eq:ocp}. 
This comes with an increased computational demand, however since we approximate the predictive controller later on, real time capability of the MPC is of minor importance.

Throughout this work, we solve \eqref{eq:ocp} using CasADi \cite{Andersson2019CASADI} and direct multiple shooting \cite{BOCK_MS}.

\section{REACHABILITY}
Reachability analysis is a key concept for our proposed method. 
Therefore, in this section we first introduce the set of closed-loop reachable states, inputs and performance deviation.
Since exact computation of these sets is challenging, we show how we can approximately compute the sets using Monte Carlo methods. 
This allows us to give probabilistic guarantees for the approximated controller.

\subsection{Reachable Sets}

Given a set of states $\mathcal{X}$ for which a solution to \eqref{eq:ocp} exists, \ie $\mathcal{X} \subseteq \mathcal{X}_\text{feas}$, we define the following $j$ step reachable set %
\begin{align} \label{eq:NstepReachSet_states_exact}
    \mathcal{R}^j_{x,\tilde{\kappa}}(\mathcal{X}) &= \{ f^{\circ j}_{\Tilde{\kappa}} (x) :  x \in \mathcal{X}\}.
\end{align}

The union of all $j$ step reachable sets is 
\begin{align}\label{eq:ReachSet_states_exact}
    \mathcal{R}_{x,\tilde{\kappa}}(\mathcal{X}) &= \bigcup_{j = 0}^{N_\text{sim}} \mathcal{R}_{x,\tilde{\kappa}}^j(\mathcal{X}),%
\end{align}
which describes the set of all closed loop reachable states under a controller  $\tilde{\kappa}$.
We use this set for multiple purposes in this work.
In a first step, we use \eqref{eq:ReachSet_states_exact} to find the closed loop relevant states of the MPC law in presence of input disturbances. 
For the design of the approximated controller, we can then restrict our investigations to this set.
That is, we can use the set to validate the desired controller specifications.
The set of possible  control inputs is given by
\begin{align} \label{eq:ReachInput}
    \mathcal{R}_{u,\Tilde{\kappa}}(\mathcal{X}) = \{ \Tilde{\kappa}(x) : x \in \mathcal{R}_{x,\tilde{\kappa}}(\mathcal{X}) \}.
\end{align}
Therefore, if $\mathcal{R}_{u,\Tilde{\kappa}} \subseteq \mathcal{U}_\text{con}$ the controller satisfies the input constraints in closed loop.
Additionally, we can compute the set of closed loop performance deviations between the MPC law $\kappa$ and an approximated controller $\Tilde{\kappa}$
\begin{align} \label{eq:reachPerformance}
    \mathcal{R}_{P,\tilde{\kappa}}(\mathcal{X}) &= \{ P_{\Tilde{\kappa},\text{abs}}(x) : x\in \mathcal{R}_{x,\tilde{\kappa}}(\mathcal{X}) \},
\end{align}
which we use to validate the performance of an approximated controller $\Tilde{\kappa}$. That is, if $\max_{x\in \mathcal{R}_{x,\tilde{\kappa}}(\mathcal{X})} P_{\tilde{\kappa},\text{abs}}(x) \leq \varepsilon_P$ for some desired threshold $\varepsilon_P$, we say that the approximated controller satisfies the performance specifications.

\subsection{Monte Carlo Method for Computation of Reachable Sets}
There are multiple methods to approximately compute reachable sets for nonlinear systems, see \eg \cite{Meyer_IntReachBook, AlthoffCora} for an overview. %
However,  most of the established methods tend to be too conservative and computationally demanding to apply. %

To address this issue, we suggest estimating the reachable sets using Monte Carlo methods as described in \cite[Chapter 7]{Meyer_IntReachBook}. %
While this way we loose deterministic guarantees that the computed sets are overapproximations of the exact reachable sets, it comes with several advantages.
First of all, using a Monte Carlo method allows us to compute the reachable set \eqref{eq:ReachSet_states_exact}, even for the case of the implicitly defined predictive controller $\kappa$ in \eqref{eq:mpclaw}, \ie for $\tilde{\kappa}(x) = \kappa(x)$.
Secondly, even if the approximated controller $\Tilde{\kappa}$ is available explicitly, we found during our work that accurately computing the reachable set of the controlled system can become computationally challenging.

For simplicity, we restrict ourselves to approximation of reachable sets using interval hulls $H$.
We define the interval hull of a set $A$ as $H(A) = \{a \in A| a_\text{lb} \leq a \leq a_\text{ub} \}$, where  $a_\text{lb}$ denotes the lower bound and $a_\text{ub}$ the upper bound of $A$.

We can provide probabilistic guarantees for the sets based on scenario optimization, see \eg  \cite{CampiScenarioApproachBook, devonport20_estreach} for details.
We summarize the results in Theorem~\ref{thm:MCMGuarantees}

\begin{theorem}[Proposition $7.2$ in \cite{Meyer_IntReachBook}] \label{thm:MCMGuarantees}
Given a number of
\begin{align}
    N_s \geq \frac{1}{\varepsilon} \left( \frac{e}{e-1} \right) \left( \log \frac{1}{\omega} + 2n_x\right)
\end{align}
samples $x_i \in \mathcal{X}$ iid. according to a probability  measure $\mathbb{P}$, %
then, for a given function $\Phi$, the interval hull
\begin{align}
        H_\Phi = H( \{ \Phi(x_i) : i= 1,\ldots, N_s \})
\end{align}
satisfies $P( H_\Phi \supseteq \Phi(x) \, \forall x \in \mathcal{X}    ) \geq 1-\varepsilon$ with confidence $1-\omega$. %
\end{theorem}
In short, Theorem~\ref{thm:MCMGuarantees} states that with confidence of at least $1-\omega$ the probability that the value $\Phi(x_i)$ of a random sample $x_i \in \mathcal{X}$ does not lie in the specified interval hull is at most $\varepsilon$.

Based on Theorem~\ref{thm:MCMGuarantees} we propose Algorithm~\ref{alg:estReachset} to estimate reachable sets.
\begin{algorithm}
\caption{Algorithm to estimate reachable sets} \label{alg:estReachset}
\begin{algorithmic}[1]
\State provide a set of initial states $\mathcal{X}_0$ described as an interval
\State provide an explicitly or implicitly defined controller $\Tilde{\kappa}$
\State collect $N_s$ samples $x_i$ iid. in $\mathcal{X}_0$ according to Theorem~\ref{thm:MCMGuarantees} 
\State compute $H_j( \{f^{\circ j}_{\tilde{\kappa}}(x_i) : i=1,\ldots N_s \})$
for all simulation steps $ j= \{ 0,\ldots, N_\text{sim}\}$
\State estimate reachable set \eqref{eq:ReachSet_states_exact} as union of all intervall hulls $\mathcal{R}_{x,\tilde{\kappa}}(\mathcal{X}_0) \approx \bigcup_j H_j$
\State collect $N_s$ samples $x_i$ iid. in $\mathcal{R}_{x,\tilde{\kappa}}(\mathcal{X}_0)$
\State estimate \eqref{eq:ReachInput} as $ \mathcal{R}_{u,\tilde{\kappa}}(\mathcal{X}_0) \approx H(\{\Tilde{\kappa}(x_i)\})$
\State estimate \eqref{eq:reachPerformance}  as $\mathcal{R}_{P,\tilde{\kappa}}(\mathcal{X}_0) \approx H(\{ 
  P_{\Tilde{\kappa},\text{abs}}(x_i)   \})$
\end{algorithmic}
\end{algorithm}

The sets constructed using Algorithm~\ref{alg:estReachset} satisfies the conditions in Theorem~\ref{thm:MCMGuarantees} and thus, with high confidence, at most an $\varepsilon$ fraction of $\mathcal{X}_0$ might not be captured within these sets. %

\section{APPROXIMATION METHOD}
In this section we present the main contribution of this work.
We first show how we can use kernel based methods to smoothly interpolate between samples from the MPC law.
Thereafter we present our proposed design algorithm, which allows to design a safe and high performing approximation of the controller. 
Since the computational complexity of the approximation scales linearly with the amount of data, we propose a function which assigns a score to a point $x$. This allows to select the most relevant data for prediction.

\subsection{Kernel-Based Interpolation}
We consider a data set $\mathcal{D}=\{ (x_i, y_i = \kappa(x_i)) \}$ of sampled states $x_i$ and corresponding control inputs $y_i = \kappa(x_i)$ for $i = \{1,\ldots, N_D \}$, where $N_D$ is the number of data points. Since the samples from the MPC law are noiseless, we try to find a smoothly interpolating function.

More formally, we express this goal as the solution to the optimization problem 
\begin{argmini} 
    {\hat{\kappa}}{\norm{\hat{\kappa}}_{\mathcal{H}}^2 }{}{\Tilde{\kappa}=}
    \addConstraint{\hat{\kappa}(x_i)}{= \kappa (x_i), \, \forall i\in\{1,\ldots,N_D\}}{.}
    \label{opti:interpolation}
\end{argmini}
That is, the approximating function should have a small norm in the reproducing kernel Hilbert space (RKHS)   $\mathcal{H}$ and must match the MPC law for the considered data points, \ie $\Tilde{\kappa} (x_i) = \kappa(x_i)$. 
Briefly outlined, a function in the RKHS can be expressed as 
$\hat{\kappa} = \sum_{j = 1}^{\infty} c_j k(\cdot,x_j)$ for some $c_j \in \mathbb{R}$, $x_j \in \mathcal{X}$ and has an associated finite norm $\norm{\hat{\kappa}}_{\mathcal{H}}^2 = \sum_{j,l}^{\infty}c_j c_lk(x_j,x_l)$, which captures the smoothness of the function with respect to the kernel $k: \mathbb{R}^{n_x} \times \mathbb{R}^{n_x} \rightarrow \mathbb{R}_{\geq 0}$.

The minimizing function $\Tilde{\kappa}$ for \eqref{opti:interpolation} is 
\begin{align} \label{eq:approx}
\Tilde{\kappa}(x) &= \sum_{i=1}^{N_D} \alpha_i k(x,x_i), %
\end{align}
where $\alpha = (K+\epsilon I)^{-1}y$ with $y_i = \kappa(x_i)$ and $K_{im} = k(x_i,x_m)$ for all selected data $ \forall i,m \in \{ 1, \ldots, N_D\} $ \cite{kanagawa2018gaussian}.
For numerical reasons, we add a regularization term $\epsilon >0$ multiplied with the identity matrix $I\in \mathbb{R}^{n_D \times n_D}$ to $K$ before computing the inverse.

Notice that \eqref{eq:approx} can also be viewed as the posterior mean of a Gaussian process with noise-free data.
For a detailed presentation on RKHSs and the connection to Gaussian processes, we refer to \cite{kanagawa2018gaussian, Liang_2020}.
We propose to use \eqref{eq:approx} as approximation of the MPC law $\kappa$.
Note that the approximated controller in  \eqref{eq:approx} has online computational complexity of $\mathcal{O}(N_D)$.

\subsection{Design Algorithm}
As outlined, the computational complexity of \eqref{eq:approx} scales linearly with the amount of selected data $N_D$.
Therefore, an efficient data selection is a primary goal in the design algorithm.
To this end, we propose to use a function which assigns a score $S:\mathbb{R}^{n_x} \rightarrow \mathbb{R}_{\geq 0}$ to a point $x$.
This function should capture, how promising it is to include a new point in the data set and is a design choice.
We propose to use a combination of different individual scores
\begin{align} \label{eq:Score}
    S(x) = \sum_j c_j S_j(x),
\end{align}
where $c_j \geq 0$ is a weighting factor for Score $S_j$.

For example, we can use the absolute and relative performance deviation and the deviation between control inputs, \ie  
$S_1(x) = P_{\Tilde{\kappa},\text{abs}}(x)$, 
$S_2(x) =  P_{\Tilde{\kappa},\text{rel}}(x)$, 
$S_3(x) = \norm{\kappa(x)-\Tilde{\kappa}(x)}$, 
to evaluate how good it is to include a new point $x$ in the data set.
However, other criteria are possible as well.  For example, one might want to add the distance to the closest point in the data set as individual score.
Note that this is a generalization to the way data are selected in \cite{RoseLearnMPCGP}, where only $S_3$ in \eqref{eq:Score} is used for selection.
To design the approximated controller we present  Algorithm~\ref{alg:GP_design}.

\begin{algorithm} %
\caption{Design of the approximated controller}\label{alg:GP_design}
\begin{algorithmic}[1]
\Require Estimate of MPC closed loop reachable set $\mathcal{R}_{x,\kappa}(\mathcal{X}_0)$, for example by using Algorithm~\ref{alg:estReachset} 
\State provide a kernel function $k:\mathbb{R}^{n_x}\times\mathbb{R}^{n_x} \rightarrow \mathbb{R}_{\geq 0} $ and a  function $S:\mathbb{R}^{n_x} \rightarrow \mathbb{R}_{\geq 0}$, see \eqref{eq:Score}
\State initialize $\mathcal{D} = \emptyset$, 
\Repeat 
\State  select sample $x_i \gets \argmaxA_{x\in \mathcal{R}_{x,\kappa}(\mathcal{X}_0)} S(x)$
\State compute control input $y_i \gets \kappa(x_i)$
\State update data $\mathcal{D}\gets \{\mathcal{D},(x_i,y_i)\}$ and corresponding controller $\Tilde{\kappa}$, see \eqref{eq:approx}

\State Safety and performance verification: Compute $\mathcal{R}_{x,\tilde{\kappa}}(\mathcal{X}_0), \, \mathcal{R}_{u,\tilde{\kappa}}(\mathcal{X}_0)$ and $\mathcal{R}_{P,\tilde{\kappa}}(\mathcal{X}_0)$ using Algorithm~\ref{alg:estReachset}.
\Until $\mathcal{R}_{x,\tilde{\kappa}}(\mathcal{X}_0) \subseteq \mathcal{X}_\text{con} \, \text{and} \, \mathcal{R}_{u,\tilde{\kappa}}(\mathcal{X}_0) \subseteq \mathcal{U}_\text{con}$ and $\max_{x\in \mathcal{R}_{x,\tilde{\kappa}}(\mathcal{X}_0)} P_{\tilde{\kappa},\text{abs}}(x) \leq \varepsilon_P$%
\end{algorithmic}
\end{algorithm}

As input to the algorithm we provide an estimate of the closed-loop reachable set   $\mathcal{R}_{x,\kappa}(\mathcal{X}_0)$. We can obtain this set by Algorithm~\ref{alg:estReachset}.
We then select a point within $\mathcal{R}_{x,\kappa}(\mathcal{X}_0)$ which maximizes the score $S(x)$ and compute the corresponding MPC input \eqref{eq:mpclaw}. 
Thereafter, we update the approximated controller  \eqref{eq:approx} by adding this point to the data set $\mathcal{D}$.
We then check if the updated approximated controller satisfies demanded specifications like constraint satisfaction and performance requirements.
If these requirements are not met we repeat the overall procedure.
Consequently, if Algorithm~\ref{alg:GP_design} terminates the approximated controller is guaranteed, in the sense of Theorem~\ref{thm:MCMGuarantees}, to satisfy all demanded specifications.

\section{CASE STUDY}
For sake of a clear and brief presentation, we illustrate our approach using a two dimensional inverted pendulum system \footnote{However, the illustrated approach is also applicable for high dimensional systems. An in-depth analysis how the approach scales with the systems dimension is subject to future research.}.
We first show, how we design the underlying optimal control problem.
Thereafter, we show how we can use our proposed method to design an approximation of the implicitly defined predictive controller.

\subsection{Inverted Pendulum}
We consider the same example of a simplified but nonlinear inverted pendulum model as in \cite{RoseLearnMPCGP} to allow a comparison between both methods.
However, in contrast to the example in \cite{RoseLearnMPCGP} we are not required to design a provably robust MPC. 
Therefore, we opt to design an MPC with only nominal stability guarantees using a discrete time version of quasi-infinite horizon MPC, see \cite{RAJHANS_DiscreteqInfty} for details. At this point, we want to mention that quasi-infinite horizon MPC is known to invoke some inherent robustness properties \cite{YU_inherentRobustness}, which we do not analyze here in detail. Consequently, it is reasonable to expect the MPC law to be robust to some small perturbations. 

\begin{figure}[t]\includegraphics{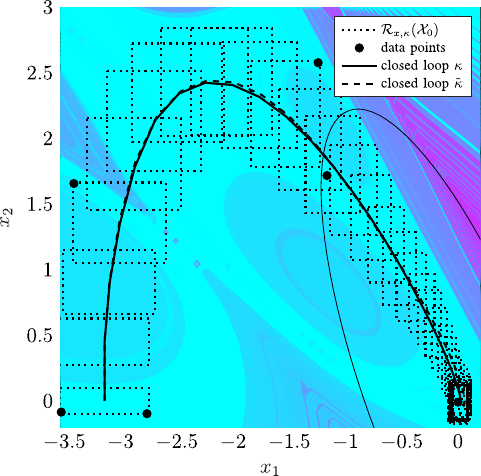}%
    \caption{Results of the proposed approach. The contour plot indicates the difference between the approximated controller and the MPC law $\norm{\kappa(x) - \Tilde{\kappa}(x)}$, where light blue indicates a small value below $0.2$ and dark blue to red indicate larger values up to $1.8$.}
    \label{fig:DeltaU_reach_cl}
    \vspace{-5mm}
\end{figure}

We choose $\ell(x,u) = x^TQx + u^TRu$, with $Q = \mathrm{diag}([10,1])$ and $R = 1$.
We linearize the system around zero and obtain a system of the form $x(k+1) = Ax(k)+Bu(k)$. Based on that we design an LQR to obtain $u = Kx$ with $K = \small \begin{bmatrix}3.85 & 2.96 \end{bmatrix}$ and we denote $A_K = (A+BK)$.
 To obtain the terminal region we first solve a discrete time Lyapunov equation $A_K^TPA_K-P=- (  Q+K^TRK  + \Delta Q)$, where we choose  $\Delta Q =\mathrm{diag}([1,1]) $ and get $P=\small \begin{pmatrix}126.89 & 50.72\\50.72 & 38.00\end{pmatrix}$.
Finally, we adjust $\alpha = 87.7$ according to method $2$ in \cite{RAJHANS_DiscreteqInfty} and use a terminal region $\mathcal{X}_\text{term} = \{x| x^TPx\leq \alpha \}$. Part of the corresponding boundary of the terminal region is outlined using a black ellipsoidal line in Fig.~\ref{fig:DeltaU_reach_cl}.
In contrast to \cite{RoseLearnMPCGP}, we now assume that the initial conditions are restricted to the set $\mathcal{X}_0 = \left[ -\pi - \frac{\pi}{8}, \,  -\pi + \frac{\pi}{8}\right] \times \left[ -0.1,\, 0.1\right]$.

To estimate the reachable set of MPC closed loop trajectories $\mathcal{R}_{x,\kappa}(\mathcal{X}_0)$, we simulate the controlled system for $N_s = 2455$ initial conditions iid in $\mathcal{X}_0$ according to Theorem~\ref{thm:MCMGuarantees} with $\varepsilon = 10^{-2}$ and $\omega = 10^{-5}$ using Algorithm~\ref{alg:estReachset}.
To account for errors that the approximated controller might make later, we apply $u = \kappa(x) \pm r$, where $r$ is a number drawn from a uniform distribution in the interval $r \in \left[-0.5, \, 0.5 \right]$, when simulating the closed loop trajectories. Consequently, we use $\tilde{\mathcal{U}} = \{ u\in\mathbb{R}^1 \, | \, \norm{u} \leq 0.5 \}$ in \eqref{eq:ocp} to tighten the input constraints.%

We depict the estimated region  $\mathcal{R}_{x,\kappa}(\mathcal{X}_0)$ in Fig.~\ref{fig:DeltaU_reach_cl} with black dashed boxes.
Notice that, due to the introduced error, we can only stay in a small region around zero.
In the next step, we use Algorithm~\ref{alg:GP_design} to design the approximated controller $\Tilde{\kappa}$, see \eqref{eq:approx}.
To this end we use a so called neural network kernel \cite{Rasmussen2006}.
We mainly decided on this kernel because it can approximate discontinuous control laws reasonably well. Moreover, in regions where data are sparse the predictions tend to stay close to the nearest data. %

To decide which data to include in \eqref{eq:approx}, we restrict our analyzes to the estimated region $\mathcal{R}_{x,\kappa}(\mathcal{X}_0)$. 
As a scoring criterion, we use \eqref{eq:Score}, with 
$c_j = \frac{1}{\max_x S_j(x)}$.
In each step of Algorithm~\ref{alg:GP_design} we update  $\Tilde{\kappa}$ by including the point which maximizes the score \eqref{eq:Score} and then analyze the resulting controller. 
In this case, we stopped the algorithm after including the $6$ points depicted as black filled circles in Fig.~\ref{fig:DeltaU_reach_cl}.
The corresponding estimated set of reachable inputs is $\mathcal{R}_{u,\tilde{\kappa}}(\mathcal{X}_0) \approx \left[ -4.78, \, 2.89 \right]$.
Notice, that this set violates the tightened input constraints in \eqref{eq:ocp}, however it is a subset of the actual input constraints, \ie $\mathcal{R}_{u,\tilde{\kappa}}(\mathcal{X}_0) \subseteq \mathcal{U}_\text{con}$.
The absolute performance error is $ \mathcal{R}_{P,\tilde{\kappa}}(\mathcal{X}_0) \approx \left[ 0, \, 3.78 \right]$ with sampling  mean $0.62$. %
 The relative performance error is in the set $\left[ 0, \, 0.61 \right]$ with sampling mean $0.006$.
In Fig.~\ref{fig:DeltaU_reach_cl}, we additionally depict the closed loop trajectory when applying the MPC law with a black line and when applying the approximated controller with a black dashed line, for $x(0) = \left[-\pi \quad  0\right]^T $. 
Furthermore, the contour plot in Fig.~\ref{fig:DeltaU_reach_cl} indicates the difference between the MPC law $\kappa$ and the approximated controller $\Tilde{\kappa}$, \ie $\norm{ \kappa(x) - \Tilde{\kappa}(x)}$. We can see that inside $\mathcal{R}_{x,\kappa}(\mathcal{X}_0)$ the difference is below $0.2$ for most states.
However, outside of this set the difference between both controller can be larger. This is not problematic as we expect these states to be unreachable in closed loop.
For improved visibility, we have not depicted the reachable set of the approximated controller $\mathcal{R}_{x,\tilde{\kappa}}(\mathcal{X}_0)$ as it is a subset of $\mathcal{R}_{x,\kappa}(\mathcal{X}_0)$, \ie $\mathcal{R}_{x,\tilde{\kappa}}(\mathcal{X}_0) \subset \mathcal{R}_{x,\kappa}(\mathcal{X}_0)$.
Finally, we summarize the computation time statistics of the MPC and the approximated controller in Tab.~\ref{tab:comptime}. Here, we see that the approximated controller is about $3000$ to $4000$ times faster than the implemented model predictive controller.

\begin{table}[!h]
\centering
\caption{Computation time statistics.}
\label{tab:comptime}
\begin{tabular}{c | c c}
  & MPC (in \si{\milli\second}) & Approximate (in \si{\micro\second})  \\
\hline
 mean & 28.8 & 7.7  \\
 median & 27.5 &  7.3 \\
 std & 19.5 &   1.4 \\
 min & 21.9 &  6.9 \\
 max &  73.4 &  17.7 \\
\end{tabular}
\end{table}

\section{CONCLUSION}

In this paper we proposed a method to approximate a model predictive controller using kernel interpolation methods. 
Since these approaches scale with the number of data which are used for the approximation, we proposed to use a scoring function to identify the most promising data points.
Additionally, we restricted our considerations to the set of closed loop reachable states. 
Depending on the possible initial conditions, this set can be small compared to the whole state space.
Overall, this led to an efficient design procedure and approximating functions with low computational demand.
Moreover, we used reachability analysis based on Monte Carlo methods to guarantee, in  probability, desired properties like safety and high performance of the approximated controller.

One advantage of the proposed design is that we can analyze the controller in each step of the proposed algorithm and decide if it satisfies our desired specifications.
In contrast,  methods that are primarily based on neural networks typically require  verification of the controller after training. 
If the approximated controller then  does not satisfy desired specifications one has to retrain the network again from scratch. 
Another key advantage of the proposed design is that we do not require the MPC to have robust stability guarantees, thereby simplifying the overall design.

In future work, we plan on applying the proposed approach to a real world system and extend the results to the trajectory tracking case.

\bibliographystyle{IEEEtran.bst}
\bibliography{Mybib}
\end{document}